\documentclass[pre,twocolumn,showpacs,aps,twoside]{revtex4}
\usepackage{graphicx}

\newcommand{\pd}[2]{\frac{\partial #1}{\partial #2}}
\newcommand{\dd}[2]{\frac{d #1}{d #2}}

\begin{document}
%\draft
\author{Colm Connaughton}
\affiliation{ Laboratoire de Physique Statistique de l'Ecole
Normale Sup\'erieur, associ\'e au CNRS, 24 Rue Lhomond, 75231
Paris Cedex 05, France}
\author{R. Rajesh}
\affiliation{Martin Fisher School of Physics, Brandeis University,  
Mailstop 057, Waltham, MA 02454-9110, USA}
\author{Oleg Zaboronski}
\affiliation{Mathematics Institute, University of Warwick, Gibbet Hill Road,
Coventry CV4 7AL, United Kingdom}
\date{\today}
\title{Stationary Kolmogorov Solutions of the Smoluchowski Aggregation
Equation with a Source Term}

\begin{abstract}
In this paper we show how the method of Zakharov transformations may be used
to analyze the stationary solutions of the Smoluchowski 
aggregation equation with a source term for arbitrary homogeneous coagulation
kernel.  The resulting power-law mass 
distributions are of Kolmogorov type in the sense that they carry a constant 
flux of mass from small masses to large. They are valid for masses much larger
than the characteristic mass of the source. We derive a ``locality criterion'',
expressed in terms of the asymptotic properties of the kernel, that must
be satisfied in order for the Kolmogorov spectrum to be an admissible
solution. Whether a given kernel leads to a gelation transition or not can be
determined
by computing the mass capacity of the Kolmogorov spectrum. As an example,
we compute the exact stationary state for the family of kernels,
$K_\zeta(m_1,m_2)=(m_1m_2)^{\zeta/2}$ which includes both gelling and
non-gelling cases, reproducing the known solution in the case $\zeta=0$.
Surprisingly, the Kolmogorov constant is the same for all kernels
in this family.
\end{abstract}

\pacs{82.20.Nk, 82.33.Ln}

\maketitle

\section{Introduction}
\label{sec-intro}
Smoluchowski's coagulation equation provides a mean field description of a 
variety of aggregation phenomena 
\cite{smoluchowski1917,chandrasekhar1943,ernst1986,aldous1999}. 
The physical picture to bear in mind is one
of a suspension of particles of varying masses that are moving around in 
$d$-dimensional space due to some transport mechanism. 
When two particles come into contact they stick
together with some probability to form a new particle whose mass is the sum of
the masses of the two constituent particles. Aggregation is irreversible
in the sense that large aggregates are not permitted to break up into 
smaller ones. If it is assumed that there are no spatial correlations between 
aggregates then the concentration of particles of mass $m$, $c(m,t)$, obeys the
Smoluchowski kinetic equation:
\begin{widetext}
\begin{eqnarray}
\nonumber  \pd{c(m,t)}{t} &=& \frac{\lambda}{2}
\int_0^{\infty}dm_1 dm_2
K(m_1,m_2,m) c(m_1,t) c(m_2,t) \delta(m-m_1-m_2)\\
\label{eq-smol}&-&\frac{\lambda}{2}
\int_0^{\infty}dm_1dm_2
K(m_1,m, m_2)c(m,t) c(m_1,t)\delta(m_2-m-m_1)\\
\nonumber&-&\frac{\lambda}{2}
\int_0^{\infty}dm_1dm_2
K(m,m_2,m_1)c(m,t) c(m_2,t)\delta(m_1-m_2-m)\\
\nonumber  
&+& \frac{J_0}{m_0}\,\delta(m-m_0) - \frac{J[c]}{M}\,\delta(m-M).
\end{eqnarray}
\end{widetext}
The kernel $K(m_1,m_2,m)$ and the constant $\lambda$ control the rate at 
which particles of masses 
$m_1$ and $m_2$ react to create particles of mass $m=m_1+m_2$. The 
$\delta(m-m_0)$ term provides a source of particles of mass $m_0$ such
that the total rate of mass input is given by $J_0$ which we take to be 
constant in time. $J[c]$ represents the mass flux which is 
functionally dependent on the entire spectrum,
$c(m,t)$. Thus the $\delta(m-M)$ term provides a sink by removing
particles from the system whose masses exceed $M$.

The kernel must be symmetric in its first two arguments, 
$K(m_1,m_2,m)=K(m_2,m_1,m)$, if it is to describe a physical aggregation 
process. Owing to the presence of the delta functions, the kernel is 
effectively a function of two arguments rather than three and is usually
written as $K(m_1,m_2)$. We include the explicit dependence on the third
argument only for notational convenience.
After writing $K(m_1,m_2,m)=K(m_1,m_2)$, some simple manipulations
reduce Eq.~(\ref{eq-smol}) to the more ``standard'' form often considered
in the literature:
%\begin{widetext}
\begin{eqnarray}
\nonumber\pd{c(m,t)}{t} &=& \frac{\lambda}{2} \int_0^\infty dm_1dm_2 
K(m_1,m_2) c(m_1,t) c(m_2,t)\\ 
& &\nonumber \hspace{-0.5cm}\left[
\delta(m-m_1-m_2)-\delta(m-m_1)-\delta(m-m_2)\right]\\
\label{eq-smol1} &+& \frac{J_0}{m_0}\,\delta(m-m_0) -\frac{J[c]}{M}\,\delta(m-M).
\end{eqnarray}
%\end{widetext}
Note that the addition of the aforementioned source and sink terms allows a 
time independent steady state
to be reached in the limit of large time. This is the main subject of this 
paper. We shall be
concerned with the situation where $m_0\to 0$ and $M\to\infty$, bearing in 
mind that the presence of a sink at infinity may be required even at finite
times in the case of the so-called ``gelling'' kernels. 
As mentioned already, we only consider here sources for which the total flux of
mass into the system, $J_0$, is constant. In the turbulence literature the
interval of masses for which $m_0 \ll m \ll M$ is called an 
{\em inertial range}.
The stationary states considered in this article are valid in this range. 

The details of the
transport mechanism and sticking probability are assumed to be built into
the kernel, $K(m_1,m_2)$, of the Smoluchowski equation. Different kernels
arise in different physical contexts and determine how the solution of the
equation should be interpreted physically. We refer to
Refs.~\cite{ernst1986,aldous1999} 
for a short
list of commonly considered kernels and their physical and/or mathematical
contexts. 

Most of the kernels of physical interest are homogeneous functions of their
arguments.  We shall denote the degree of homogeneity of the kernel by
$\zeta$. That is
\begin{equation}
K(hm_1,hm_2,hm) = h^\zeta K(m_1,m_2,m).
\end{equation}
This homogeneity need not be uniformly weighted between the two arguments.
Following \cite{ernst1986}, we introduce exponents, $\mu$ and $\nu$ to
take into account this fact:
\begin{equation}
K(m_1,m_2,m) \sim m_1^\mu m_2^\nu \ \ \ \mbox{for $m_2 \gg
m_1$.}
\end{equation}
The exponents $\mu$ and $\nu$ satisfy $\mu + \nu=\zeta$. Let us consider a 
couple of simple examples to clarify our notation. The kernel 
$K(m_1,m_2) = \lambda(m_1^{1+\epsilon} + m_2^{1+\epsilon})$ has $\zeta=1+\epsilon$, $\mu=0$ and $\nu=1+\epsilon$ whereas the kernel 
$K(m_1,m_2) = \lambda(m_1^{1/3}+m_2^{1/3})(m_1^{-1/3}+m_2^{-1/3})$ has 
$\lambda=0$, $\mu=-1/3$ and $\nu=1/3$.
These basic properties of the kernel are all we shall require for what follows.

In this paper, we study the steady state behavior of $c(m)$ when $m_0 \ll m
\ll M$. In Sec.~\ref{sec-dimensional}, using dimensional analysis, we derive
the large mass dependence of $c(m)$. It is also shown that the power law
spectrum corresponds to a constant flux of mass in mass space.
In Sec.~\ref{sec-ZakharovTransformations}, we show that the Smoluchowski
equation is mathematically very similar to the kinetic equation for 3-wave
turbulence. Using Zakharov transformations from 3-wave turbulence, we
rederive the mass spectrum as well as compute the amplitude also known as the
Kolmogorov constant. 
The characteristic mass of the source, $m_0$, does not appear in the
dimensional argument.
In Sec.~\ref{sec-locality}, we find the conditions under which this
assertion is correct when we address the
question of the locality of the mass cascade.
In Sec.~\ref{sec-capacity}, we discuss the notion of mass capacity of the 
Kolmogorov
spectrum and show how it distinguishes between gelling and non-gelling
kernels. In Sec.~\ref{sec-example}, we explicitly compute the Kolmogorov
spectrum for a one-parameter family of kernels given by 
$K_\zeta(m_1,m_2)=(m_1m_2)^{\zeta/2}$. We find that the value of the
Kolmogorov constant is the same for all models in this family.
Finally, we end with a summary in Sec.~\ref{sec-summary}.

\section{Dimensional Derivation of the Stationary Spectrum
\label{sec-dimensional}}

Before proceeding into detailed analysis of the stationary states of model
Eq.~(\ref{eq-smol}) let us first describe intuitively what we mean by a 
Kolmogorov
solution by employing a simple scaling argument. We shall use the simplified
form Eq.~(\ref{eq-smol1}) for brevity. The stationary energy distribution of forced
hydrodynamic turbulence is described by the famous Kolmogorov $5/3$ spectrum
(for instance, see \cite{frischBook}).
This spectrum, postulated from dimensional considerations, carries a constant
flux of energy from large scales to small by means of 
vortex-vortex interactions.
The analogous cascade for the Smoluchowski equation is a cascade of mass
from small particles to large mediated by the coagulation of aggregates. The 
Kolmogorov spectrum for aggregation carries a constant flux of mass.

The physical
dimensions of the various quantities appearing in Eq.~(\ref{eq-smol1})
are as follows : $\left[c\right] = {\rm M}^{-1}\,{\rm L}^{-d}$, 
$\left[J\right] = {\rm M}\,{\rm L}^{-d}\,{\rm T}^{-1}$ and 
$\left[\lambda \right] = {\rm M}^{-\zeta}\,{\rm L}^{d}\,{\rm T}^{-1}$. 
If we now take the 
combination $c \sim J^\gamma\lambda^\alpha m^\beta$, simple dimensional 
analysis requires that we choose $\gamma=1/2$, $\alpha=-1/2$ and $\beta=-(\zeta+3)/2$. 
Dimensional considerations therefore lead us to a spectrum of the form
\begin{equation}
\label{eq-KExponent}
c(m) \sim \sqrt{\frac{J_0}{\lambda}}\,m^{-\frac{\zeta+3}{2}}.
\end{equation}

The characteristic mass of the source, $m_0$, does not appear in our
dimensional argument on the basis that we expect this solution to be valid
for masses much greater than $m_0$. We find the conditions under which this
assertion is correct in Sec.~\ref{sec-locality} when we address the
question of the locality of the mass cascade.
                                                                                
The exponent Eq.~(\ref{eq-KExponent}) is not new. It appeared in early work by
Hendriks, Ernst and Ziff \cite{hendriks1983} as the scaling of the post-gel
stage of gelling systems. Their work makes an implicit connection between
this scaling and the fact that there is a mass flux out of the system in the
post-gel stage. It was then derived explicitly for the Smoluchowski equation
with source term by Hayakawa \cite{hayakawa1987} for a particular family of
kernels but without making any connection with the physical role played by
the mass flux.

That the spectrum Eq.~(\ref{eq-KExponent}) corresponds to a constant flux of mass in mass space 
is easily seen from the following scaling
argument. We express the local conservation of mass by means of the 
continuity equation
\begin{equation}
\label{eq-consLaw}
\pd{m c(m,t)}{t} = -\pd{J(m,t)}{m},
\end{equation}
where
\begin{eqnarray}
&&\pd{J(m,t)}{m} =- \frac{m\lambda}{2} \int_0^\infty\!\!\!\! dm_1dm_2 
\big[K(m_1,m_2) c(m_1,t) 
\times \nonumber \\
&& c(m_2,t) 
\left\{\delta(m\!-\!m_1\!-\!m_2)\!-\!\delta(m\!-\!m_1)\!-\!\delta(m\!-\!m_2)
\right\}\big]. 
\label{eq-flux1}
\end{eqnarray}
We now assume a stationary spectrum, $c(m) = C m^{-x}$. By introducing 
new variables, $m_1=m \mu_1$, $m_2=m \mu_2$ and using the scaling properties
of the kernel we deduce that
\begin{equation}
\pd{J}{m} \propto m^{2+\zeta-2 x}
\end{equation}
with the constant of proportionality being given by the integral expression
which remains after scaling out the $m$ dependence of the RHS of Eq.~(\ref{eq-flux1}).
Thus
\begin{equation}
J(m)\propto m^{3+\zeta-2x}.
\end{equation}
It is clear from Eq.~(\ref{eq-consLaw}) that in order to have a stationary state,
$J$ must be independent of $m$ which determines the exponent of the Kolmogorov
spectrum as
\begin{equation}
\label{eq-Kexponent}
x_K=\frac{3+\zeta}{2}.
\end{equation}
Of course we cannot determine the Kolmogorov constant, $C$, from such a 
scaling argument. In addition the validity of our scaling argument depends on
the convergence of the various integral expressions which have been hidden
behind proportionality signs. 

In this paper we address these short-comings by computing the exact stationary
solutions of Eq.~(\ref{eq-smol1}) using the method of Zakharov 
transformations borrowed from the theory of wave turbulence. We obtain
the exponent, $x_K$ expected from scaling considerations and the value of
the Kolmogorov constant, $C$. An answer is obtained for arbitrary homogeneous
kernels. However the analysis involves the exchange of orders of integration
on the RHS of Eq.~(\ref{eq-smol1}). It is thus necessary to check a posteriori that
the RHS is convergent on the prospective spectrum in order that it be an
admissible solution. This check leads to a ``locality criterion'', namely
\begin{equation}
\label{eq-locality}
\mu-\nu+1 >0,
\end{equation}
which must be satisfied by the kernel in order that the Kolmogorov spectrum
be realizable. 
%We discuss also the notion of mass capacity of the Kolmogorov
%spectrum and show how it distinguishes between gelling and non-gelling
%kernels. Finally, in the last section we explicitly compute the Kolmogorov
%spectrum for a one-parameter family of kernels given by 
%$K_\zeta(m_1,m_2)=(m_1m_2)^{\zeta/2}$. We find that the value of the
%Kolmogorov constant is the same for all models in this family.

\section{Zakharov Transformation for Smoluchowski Equation}
\label{sec-ZakharovTransformations}
To find the stationary solutions of Eq.~(\ref{eq-smol}) in the situation
$m \gg m_0$, $J_0 =$ constant, we must solve:
\begin{widetext}
\begin{eqnarray}
0 &=& \!\!\frac{\lambda}{2}\!\!
\int_0^{\infty}\!\!\!\!\!\!dm_1dm_2\,
K(m_1,m_2,m)\,c(m_1)c(m_2)\,\delta(m\!-\!m_1\!-\!m_2)
%\nonumber \\
%&-&
-\!\frac{\lambda}{2}\!\!
\int_0^{\infty}\!\!\!\!\!\!dm_1dm_2\,
K(m_1,m,m_2)\,c(m)c(m_1)\,\delta(m_2\!-\!m\!-\!m_1)\nonumber \\
&-&\!\!\frac{\lambda}{2}\!\!
\int_0^{\infty}\!\!\!\!\!\!dm_1dm_2\,
K(m,m_2,m_1)\,c(m)c(m_2)\,\delta(m_1\!-\!m_2\!-\!m).
\label{se_integral}
\end{eqnarray}
\end{widetext}
Structurally this equation is very similar to the kinetic equation for wave
turbulence with a 3-wave interaction. For an introduction to the theory of wave
turbulence see \cite{zakharovBook}. A useful trick for finding the stationary 
power law solutions of such equations was devised by Zakharov 
\cite{zakharov1967, zakharov1966} in the late 60's and is easily applied here.
Restricting ourselves to power law solutions of the form $c(m) = C m^{-x}$,
we apply the following changes of variables,
\begin{eqnarray}
\label{Z1}(m_1,m_2) &\to& (\frac{m m_1^\prime}{m_2^\prime},\frac{m^2}{m_2^\prime})\\
\label{Z2}(m_1,m_2) &\to& (\frac{m^2}{m_1^\prime},\frac{m m_2^\prime}{m_1^\prime}).
\end{eqnarray}
to the second and third integrals in Eq.~(\ref{se_integral})
respectively. 
Dropping the primes on the transformed variables and using the homogeneity
and symmetry properties of the kernel we obtain the equation
\begin{eqnarray}
0&=&\!\!\frac{\lambda C^2}{2}\!\!\int_0^\infty\!\!\!\!\!\!\! 
dm_1dm_2 \big[K(m_1,m_2,m)
(m_1m_2)^{-x}m^{2-\zeta-2x}\! \times \nonumber \\
&&\!\!\!\left(m^{2x-\zeta-2}\!-\!m_1^{2x-\zeta-2}\!-\!m_2^{2x-\zeta-2}\right)
\!\delta(m\!-\!m_1\!-\!m_2)\big].
\end{eqnarray}
It is immediately evident that the integrand is identically zero for 
$2x-\zeta-2=1$ from which we get the same Kolmogorov exponent, 
\begin{equation}
\label{eq-Kexponent2}
x_K=\frac{3+\zeta}{2},
\end{equation}
obtained in Sec.~\ref{sec-intro} by a scaling argument. The value of the 
Kolmogorov constant can be determined by considering the local mass
flux defined from Eqs.~(\ref{eq-consLaw}) and (\ref{eq-flux1}). Restricting 
ourselves to spectra of the form $c(m)=Cm^{-x}$, the Zakharov transformation
allows us to write Eq.~(\ref{eq-flux1}) in the form
\begin{widetext}
\begin{eqnarray*}
\pd{J(m,t)}{m}&=&\!\!- \frac{\lambda m C^2}{2} \int_0^\infty \!\!
\!\!\!\!\!dm_1dm_2 K(m_1,m_2,m) (m_1m_2)^{-x}m^{2-\zeta-2x}
\!\left(m^{2x-\zeta-2}-m_1^{2x-\zeta-2}-m_2^{2x-\zeta-2}\!\right)\!
\delta(m-m_1-m_2)\\
&=& \lambda C^2 m^{2+\zeta-2x} I(x)
\end{eqnarray*}
\end{widetext}
where
\begin{eqnarray}
I(x)&=&-\frac{1}{2}\int_0^\infty d\mu_1d\mu_2 \big[K(\mu_1,\mu_2,1)
(\mu_1\mu_2)^{-x} \times \nonumber \\
&&\left(1-\mu_1^{2x-\zeta-2}-\mu_2^{2x-\zeta-2}\right)\delta(1\!-\!\mu_1\!-
\!\mu_2)
\big].
\label{eq-I}
\end{eqnarray}
From this we deduce that the flux is given by
\begin{equation}
J(m)=\frac{\lambda C^2 I(x)}{3+\zeta-2x}m^{3+\zeta-2x}.
\end{equation}
In the steady state, $x=x_K=(3+\zeta)/2$ and  the flux must be a constant 
equal to $J_0$. Thus we have
\begin{equation}
J_0=\lim_{x\to x_K} \frac{\lambda C^2 I(x)}{3+\zeta-2x}m^{3+\zeta-2x}.
\end{equation}
We know that $I(x_K)=0$ so we must apply l'Hopital's rule to evaluate the limit
to arrive at
\begin{equation}
J_0=\frac{\lambda}{2} C^2 \left.\dd{I}{x}\right|_{x_K},
\end{equation}
and hence
\begin{equation}
\label{eq-KolConstant}
C=\sqrt{\frac{2 J_0}{\lambda} \left.\dd{I}{x}\right|_{x_K}^{-1}}.
\end{equation}
The Kolmogorov solution is therefore,
\begin{equation}
\label{eq-Ksolution}
c(m) = C\,m^{-x_K},
\end{equation}
with $C$ given by Eq.~(\ref{eq-KolConstant}) and $x_K$ given by 
Eq.~(\ref{eq-Kexponent2}).

\section{Locality of the Mass Cascade \label{sec-locality}}

In the analysis of the previous section we have freely split the integrand 
on the RHS of Eq.~(\ref{eq-smol}) and exchanged orders of integration to derive 
the Kolmogorov spectrum. In order to justify these manipulations we must 
demonstrate a posteriori that the original collision integral is convergent
on the Kolmogorov spectrum. We do this in this section.
\begin{figure}
%\begin{center}
%\epsfig{file=bottle2.ps,width=10cm, angle=270}
\includegraphics[width=\columnwidth]{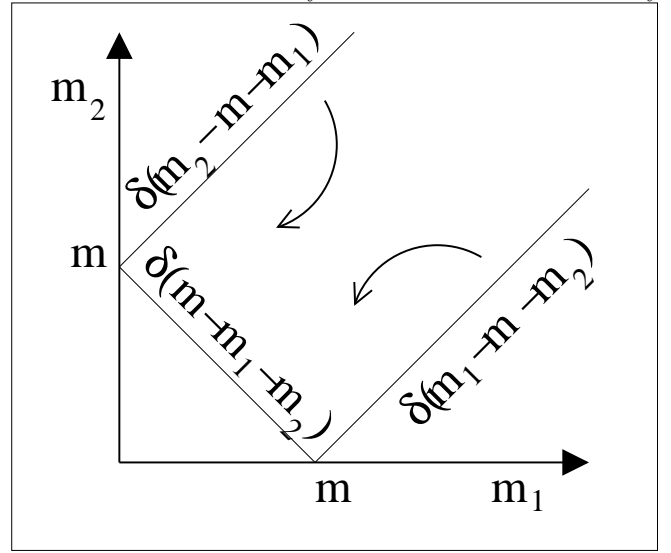}
%\end{center}
\caption{\label{fig-support}Support of the integrand of Eq.~(\ref{eq-smol})}
\end{figure}

The support of the integrand on the RHS of Eq.~(\ref{eq-smol}) is shown in 
Fig.~\ref{fig-support}. 
Since the kernel and the mass distributions which we study
are scale invariant the only possible sources of divergences are at infinity 
and at the two points $(0,m)$ and $(m,0)$ where the contour of integration
intersects the axes. Let us now study carefully the behavior of the integrand
near these points for power law mass distributions.

The behavior at infinity is easy. As $m_2\to\infty$ along the lower contour
the integrand looks like
\begin{equation}
K(m,m_2,m+m_2) c(m) c(m_2) \sim m^{\mu-x} m_2^{\nu -x}.
\end{equation}
The integral is therefore convergent as $m_2\to\infty$ if 
\begin{equation}
\label{eq-localityCriterionAtInfinity}
\nu-x<-1 \Rightarrow x > \nu+1.
\end{equation} 
The same criterion is obtained along the upper contour. 
The convergence near zero requires a little care. Near $m_1=0$ the integrand 
looks like
\begin{widetext}
\begin{eqnarray*}
& &c(m_1)\left[ K(m_1,m-m_1,m)c(m-m_1) - K(m_1,m,m+m_1) c(m)\right] \\
&=& c(m_1) \left[m_1 \left(\pd{}{\xi}\left.\left(K(m_1,\xi,m)c(\xi)\right)\right|_{\xi=m} - c(m) \pd{}{\xi}\left.\left(K(m_1,m, \xi)\right)\right|_{\xi=m}\right) + {\rm o}(m_1^2)\right]\\
&\sim& m_1^{1-x+\mu}.
\end{eqnarray*}
\end{widetext}
Note the cancellation of the leading order terms in the Taylor expansion on
the second line above.  The corresponding integral is convergent as $m_1\to 0$ 
for
\begin{equation}
\label{eq-localityCriterionAtZero}
\mu+1-x>-1\Rightarrow x<\mu+2.
\end{equation} 
The same criterion is obtained if we look near $m_2=0$. We conclude from 
Eqs.~(\ref{eq-localityCriterionAtInfinity}) and 
(\ref{eq-localityCriterionAtZero}) that a power law mass distribution, 
$Cm^{-x}$, yields a convergent
collision integral if $x$ lies in the interval $\left[\nu+1,\mu+2\right]$. The
existence of such an interval of convergence puts a constraint on the 
asymptotic behavior of the kernel, namely
\begin{equation}
\label{eq-localityCriterion}
\mu-\nu+1>0.
\end{equation}
We now must address the question of when the Kolmogorov spectrum derived in
Sec.~\ref{sec-ZakharovTransformations} lies in this interval of convergence.
The answer is surprisingly simple. Remembering that $\mu+\nu=\zeta$, it is 
immediately
evident from Eq.~(\ref{eq-Kexponent2}) that the Kolmogorov spectrum lies midway
between the two constraints Eqs.~(\ref{eq-localityCriterionAtInfinity}) and 
(\ref{eq-localityCriterionAtZero}). Therefore {\em if} an interval of
convergence exists for a given kernel, the corresponding Kolmogorov spectrum 
is an admissible stationary solution of Eq.~(\ref{eq-smol}) and it
lies at the midpoint of the interval of convergence. 

We call 
Eq.~(\ref{eq-localityCriterion}) a {\em locality} criterion since systems for which it
is satisfied can be characterized in the stationary state by a local mass flux,
$J_0$. When the spectrum is local, the details of how we take the limits
$m_0\to 0$ and $M\to\infty$ to produce a large inertial range are 
inconsequential since all integrals converge. If 
Eq.~(\ref{eq-localityCriterion}) is not satisfied then presumably the
final stationary  state depends on the details of the source/sink and is 
therefore non-universal.

We note that the kernel $m_1^{1+\epsilon}+m_2^{1+\epsilon}$, 
mentioned in the introduction, is 
marginal in the sense that it violates the locality criterion for any finite
$\epsilon$. It has been shown\cite{carr1992} that this kernel undergoes 
instantaneous gelation so perhaps there is some connection between this
phenomenon and the concept of locality. In addition, the generalised sum 
kernel, $K(m_1, m_2) = m_1^{-\mu} + m_2^{-\mu}$, which violates the
locality condition for $\mu\geq 1$, was studied extensively
by Krapivsky et al. \cite{krapivsky1998,krapivsky1999}. They found that in this case, the
system does not reach a steady state but rather continues to evolve very
slowly on a logarithmic timescale for all time.

In closing this section it should be noted that a rigorous understanding of 
the conditions under which the stationary state depends only on the local flux 
is one of the missing pieces in the theory of hydrodynamic turbulence.

\section{\label{sec-capacity}Finite and Infinite Capacity Cases -- 
Gelling and Non-gelling Kernels}

It was found in the 60's \cite{mcleod1962} that the solution of Eq.~(\ref{eq-smol1})
for certain kernels violates mass conservation within a finite time, $t^*$. 
When this violation occurs, $\lim_{m\to\infty} P(m)$ becomes finite. In the
late 70's it was found that meaningful solutions exist post-$t^*$ 
and the violation of mass conservation was given a physical interpretation in 
terms of what is now termed a ``gelation transition''
\cite{leyvraz1981, lushnikov1978, ziff1980}. Gelation occurs when there is a finite
flow of mass to an infinite mass cluster (``gel''). As a consequence, the 
total mass of the normal (``sol'') particles is no longer conserved. In order
to avoid inconsistencies, the gel particles must be considered as those 
clusters whose mass diverges as the size of a finite system is taken to 
infinity. It is now well known \cite{ernst1986} that the solutions of 
Eq.~(\ref{eq-smol1}) undergo gelation for kernels having $\zeta>1$. 

The gelation criterion, $\zeta>1$, can be given a very simple physical 
interpretation by examining the mass capacity of the Kolmogorov spectrum. If
we continue to add mass to the system at a constant rate $J_0$ then we know 
that the final steady state is given by the Kolmogorov spectrum
Eq.~(\ref{eq-Ksolution}). When the total mass contained in this solution is finite
then mass  conservation  must be violated at some time since the total mass 
supplied to the system grows linearly in time. The total mass capacity of
the Kolmogorov spectrum is finite when
\begin{displaymath}
\int_{m_0}^\infty dm\,m\, C\,m^{-\frac{3+\zeta}{2}} < \infty.
\end{displaymath} 
This integral in convergent at its upper limit when $1-(3+\zeta)/2 <-1$ or
$\zeta>1$. Thus gelation can be seen as a kind of safety valve which 
allows mass to flow out of the system when the Kolmogorov spectrum is 
incapable of absorbing all of the mass supplied to the system. Conversely, 
one would expect intuitively that infinite capacity systems should not
exhibit gelation.

\section{\label{sec-example}Example : the Family of Kernels, 
$K_\zeta(m_1,m_2,m)= (m_1m_2)^{\zeta/2}$}

In this section we explicitly evaluate the Kolmogorov constant, $C$, for the 
family of kernels, $K_\zeta(m_1,m_2,m)=(m_1m_2)^{\zeta/2}$. These kernels 
have $\mu=\nu=\zeta/2$ so that the corresponding Kolmogorov spectrum always
satisfies the locality criterion, Eq.~(\ref{eq-localityCriterion}). The family 
includes both gelling and non-gelling kernels. In general to compute, $C$, we
need to evaluate the following integral at $x=(3+\zeta)/2$ :
\begin{widetext}
\begin{eqnarray}
\label{eq-messyIntegral}
\nonumber \dd{I}{x} &=& \frac{1}{2}\int_0^1 d\mu_1 K(\mu_1,1-\mu_1,1) \left\{(\mu_1(1-\mu_1))^{-x}\right.\left[-2\mu_1^{2x-\zeta-2}\log\mu_1-2(1-\mu_1)^{2x-\zeta-2}\log(1-\mu_1)\right.\\
\label{eq-messyInt}& &\left.\left.-\log(\mu_1(1-\mu_1))(1-\mu_1^{2x-\zeta-2}-(1-\mu_1)^{2x-\zeta-2})\right]\right\}.
\end{eqnarray}
\end{widetext}
This is obtained from Eq.~(\ref{eq-I}) by integrating out $\mu_2$ and 
differentiating with respect to $x$. When we set 
$K(\mu_1,\mu_2,1)=(\mu_1\mu_2)^{\zeta/2}$ and $x=(3+\zeta)/2$ in this 
expression we find, rather surprisingly, that all dependence on $\zeta$
cancels out and we are left with
\begin{eqnarray}
\left.\dd{I}{x}\right|_{(3+\zeta)/2}&=&
-\int_0^1d\mu_1 
\frac{\mu_1\log\mu_1+(1-\mu_1)\log(1-\mu_1)}
{\mu_1^{3/2}(1-\mu_1)^{3/2}}
\nonumber \\
&=&4\pi \hspace{1cm}\mbox{\em (Mathematica)}
\end{eqnarray}
Hence the Kolmogorov solution for all kernels in this family is
\begin{equation}
c(m) = \sqrt{\frac{J_0}{2\pi\lambda}}\, m^{-\frac{3+\zeta}{2}}.
\end{equation}
To close we note that we can check our answer independently for at least 
one case. For the constant kernel with zero initial concentration, an exact 
solution of Eq.~(\ref{eq-smol1}) has been known for some time. The 
details can be found in \cite{leyvraz2003}. This solution is
\begin{equation}
c_m(t) = \sum_{k=1}^\infty c_k(t)\,\delta(m-km_0)
\end{equation}
with
\begin{equation}
c_k(t) = \frac{m_0\pi^2}{\lambda^2 J_0 t^3}
\sum_{j=-\infty}^{j=\infty}\!(2j+1)^2\left[1+
\frac{(2j+1)^2m_0\pi^2}{2\lambda J_0t^2}\right]^{-k-1}.
\end{equation}
The $t\to\infty$ limit of this expression exists can be calculated by replacing
the sum by an integral in the limit of large $t$. This integral can be 
expressed in terms of gamma functions. One finds
 \begin{eqnarray*}
\lim_{t\to \infty} c_k(t)& = & 
\sqrt{\frac{J_0}{2\pi\lambda m_0}}\frac{\Gamma(k-\frac{1}{2})}
{\Gamma(k+1)} \\
&\sim& 
\sqrt{\frac{J_0}{2\pi\lambda m_0}} k^{-\frac{3}{2}}~~\mbox{for $k \gg 1$.}
\end{eqnarray*}
Setting $m_0=1$ we recover the result of our earlier computation of the 
Kolmogorov spectrum for $m\gg1$.

As pointed out to us by one of our referees, the constant, $C$, has also been
computed \cite{krapivsky1998,krapivsky1999}  for the generalised sum kernel
\begin{equation}
K(m_1, m_2) = m_1^{-\zeta} + m_2^{-\zeta}.
\end{equation}
We computed the integral (\ref{eq-messyIntegral}) for this kernel using 
{\em Mathematica} and found the Kolmogorov constant to be
\begin{equation}
C = \sqrt{\frac{J_0 (1-\zeta^2) \cos \frac{\pi \zeta}{2}}{4\lambda\pi}} 
\end{equation}
as found in \cite{krapivsky1998,krapivsky1999} using completely different methods.

\section{\label{sec-summary}Conclusion}

To summarize, we have shown how the notion of a mass cascade analogous
to the Kolmogorov energy cascade of hydrodynamic turbulence is relevant to 
understanding the stationary state of the Smoluchowski equation with constant
mass production term. Furthermore we have shown how the exact stationary 
spectrum may be computed using the method of Zakharov transformations
and given some criteria for assessing the physical validity of this solution.
We have not made any attempt to address the important question of the validity
of the Smoluchowski equation itself in describing the statistics of 
particular aggregation models. The mean field assumption leading to 
Eq.~(\ref{eq-smol}) can be violated in several ways as discussed in 
\cite{leyvraz2003}. Of particular relevance to lattice aggregation models is 
the case where fluctuations dominate the statistics and invalidate the
mean field Smoluchowski equation \cite{zaboronski2001,krishnamurthy2002}. 
In a future publication
\cite{cardy2004} we shall address this issue for the particular case of
constant kernel stochastic aggregation where the presence of fluctuations
leads to a renormalization of the constant $\lambda$. The techniques
developed in this paper
will allow us to find the renormalized Kolmogorov spectrum as the 
stationary solution of a modified Smoluchowski equation.

\section*{Acknowlegements}
We would like to thank W. Wagner and F. Leyvraz for helpful discussions
and feedback. C.C. acknowledges financial support from Marie-Curie Fellowship
HPMF-CT-2002-02004.

\section*{Note added in proof}
We have found that V.M. Kontorovich has recently applied the Zakharov
Transformation to aggregation problems \cite{kontorovich2001} for a class
of kernels arising from the study of galactic mergers in astrophysics.

%\bibliography{ref}

\begin{thebibliography}{22}
\expandafter\ifx\csname natexlab\endcsname\relax\def\natexlab#1{#1}\fi
\expandafter\ifx\csname bibnamefont\endcsname\relax
  \def\bibnamefont#1{#1}\fi
\expandafter\ifx\csname bibfnamefont\endcsname\relax
  \def\bibfnamefont#1{#1}\fi
\expandafter\ifx\csname citenamefont\endcsname\relax
  \def\citenamefont#1{#1}\fi
\expandafter\ifx\csname url\endcsname\relax
  \def\url#1{\texttt{#1}}\fi
\expandafter\ifx\csname urlprefix\endcsname\relax\def\urlprefix{URL }\fi
\providecommand{\bibinfo}[2]{#2}
\providecommand{\eprint}[2][]{\url{#2}}

\bibitem[{\citenamefont{Smoluchowski}(1917)}]{smoluchowski1917}
\bibinfo{author}{\bibfnamefont{M.}~\bibnamefont{Smoluchowski}},
  \bibinfo{journal}{Z. Phys. Chem.} \textbf{\bibinfo{volume}{92}},
  \bibinfo{pages}{215} (\bibinfo{year}{1917}).

\bibitem[{\citenamefont{Chandrasekhar}(1943)}]{chandrasekhar1943}
\bibinfo{author}{\bibfnamefont{S.}~\bibnamefont{Chandrasekhar}},
  \bibinfo{journal}{Rev. Mod. Phys.} \textbf{\bibinfo{volume}{15}},
  \bibinfo{pages}{1} (\bibinfo{year}{1943}).

\bibitem[{\citenamefont{Ernst}(1986)}]{ernst1986}
\bibinfo{author}{\bibfnamefont{M.}~\bibnamefont{Ernst}}, in
  \emph{\bibinfo{booktitle}{Fractals in Physics}}, edited by
  \bibinfo{editor}{\bibfnamefont{L.}~\bibnamefont{Pietronero}}
  \bibnamefont{and} \bibinfo{editor}{\bibfnamefont{E.}~\bibnamefont{Tosatti}}
  (\bibinfo{publisher}{North Holland}, \bibinfo{address}{Amsterdam},
  \bibinfo{year}{1986}), p. \bibinfo{pages}{289}.

\bibitem[{\citenamefont{Aldous}(1999)}]{aldous1999}
\bibinfo{author}{\bibfnamefont{D.}~\bibnamefont{Aldous}},
  \bibinfo{journal}{Bernoulli} \textbf{\bibinfo{volume}{5}}, \bibinfo{pages}{3}
  (\bibinfo{year}{1999}).

\bibitem[{\citenamefont{Frisch}(1995)}]{frischBook}
\bibinfo{author}{\bibfnamefont{U.}~\bibnamefont{Frisch}},
  \emph{\bibinfo{title}{Turbulence: The Legacy of A. N. Kolmogorov}}
  (\bibinfo{publisher}{Cambridge University Press},
  \bibinfo{address}{Cambridge}, \bibinfo{year}{1995}).

\bibitem[{\citenamefont{Hendriks et~al.}(1983)\citenamefont{Hendriks, Ernst,
  and Ziff}}]{hendriks1983}
\bibinfo{author}{\bibfnamefont{E.}~\bibnamefont{Hendriks}},
  \bibinfo{author}{\bibfnamefont{M.}~\bibnamefont{Ernst}}, \bibnamefont{and}
  \bibinfo{author}{\bibfnamefont{R.}~\bibnamefont{Ziff}}, \bibinfo{journal}{J.
  Stat. Phys.} \textbf{\bibinfo{volume}{31}}, \bibinfo{pages}{519}
  (\bibinfo{year}{1983}).

\bibitem[{\citenamefont{Hayakawa}(1987)}]{hayakawa1987}
\bibinfo{author}{\bibfnamefont{H.}~\bibnamefont{Hayakawa}},
  \bibinfo{journal}{J. Phys. A} \textbf{\bibinfo{volume}{20}},
  \bibinfo{pages}{L801} (\bibinfo{year}{1987}).

\bibitem[{\citenamefont{Zakharov et~al.}(1992)\citenamefont{Zakharov, Lvov, and
  Falkovich}}]{zakharovBook}
\bibinfo{author}{\bibfnamefont{V.}~\bibnamefont{Zakharov}},
  \bibinfo{author}{\bibfnamefont{V.}~\bibnamefont{Lvov}}, \bibnamefont{and}
  \bibinfo{author}{\bibfnamefont{G.}~\bibnamefont{Falkovich}},
  \emph{\bibinfo{title}{Kolmogorov Spectra of Turbulence}}
  (\bibinfo{publisher}{Springer-Verlag}, \bibinfo{address}{Berlin},
  \bibinfo{year}{1992}).

\bibitem[{\citenamefont{Zakharov and Filonenko}(1967)}]{zakharov1967}
\bibinfo{author}{\bibfnamefont{V.}~\bibnamefont{Zakharov}} \bibnamefont{and}
  \bibinfo{author}{\bibfnamefont{N.}~\bibnamefont{Filonenko}},
  \bibinfo{journal}{Zh. Prikl. Mekh. Tekhn. Fiz.} \textbf{\bibinfo{volume}{6}},
  \bibinfo{pages}{62} (\bibinfo{year}{1967}).

\bibitem[{\citenamefont{Zakharov and Filonenko}(1966)}]{zakharov1966}
\bibinfo{author}{\bibfnamefont{V.}~\bibnamefont{Zakharov}} \bibnamefont{and}
  \bibinfo{author}{\bibfnamefont{N.}~\bibnamefont{Filonenko}},
  \bibinfo{journal}{Doklady Akad. Nauk. SSSR} \textbf{\bibinfo{volume}{170}},
  \bibinfo{pages}{1292} (\bibinfo{year}{1966}).

\bibitem[{\citenamefont{Carr and Costa}(1992)}]{carr1992}
\bibinfo{author}{\bibfnamefont{J.}~\bibnamefont{Carr}} \bibnamefont{and}
  \bibinfo{author}{\bibfnamefont{F.}~\bibnamefont{Costa}}, \bibinfo{journal}{Z.
  Ang. Math. Phys.} \textbf{\bibinfo{volume}{43}}, \bibinfo{pages}{974}
  (\bibinfo{year}{1992}).

\bibitem[{\citenamefont{Krapivsky et~al.}(1998)\citenamefont{Krapivsky, Mendes,
  and Redner}}]{krapivsky1998}
\bibinfo{author}{\bibfnamefont{P.}~\bibnamefont{Krapivsky}},
  \bibinfo{author}{\bibfnamefont{J.}~\bibnamefont{Mendes}}, \bibnamefont{and}
  \bibinfo{author}{\bibfnamefont{S.}~\bibnamefont{Redner}},
  \bibinfo{journal}{Eur. Phys. J. B} \textbf{\bibinfo{volume}{4}},
  \bibinfo{pages}{401} (\bibinfo{year}{1998}).

\bibitem[{\citenamefont{Krapivsky et~al.}(1999)\citenamefont{Krapivsky, Mendes,
  and Redner}}]{krapivsky1999}
\bibinfo{author}{\bibfnamefont{P.}~\bibnamefont{Krapivsky}},
  \bibinfo{author}{\bibfnamefont{J.}~\bibnamefont{Mendes}}, \bibnamefont{and}
  \bibinfo{author}{\bibfnamefont{S.}~\bibnamefont{Redner}},
  \bibinfo{journal}{Phys. Rev. B} \textbf{\bibinfo{volume}{59}},
  \bibinfo{pages}{15950} (\bibinfo{year}{1999}).

\bibitem[{\citenamefont{Mc~Leod}(1962)}]{mcleod1962}
\bibinfo{author}{\bibfnamefont{J.}~\bibnamefont{Mc~Leod}},
  \bibinfo{journal}{Quart. J. Math. Oxford Ser. (2)}
  \textbf{\bibinfo{volume}{13}}, \bibinfo{pages}{119} (\bibinfo{year}{1962}).

\bibitem[{\citenamefont{Leyvraz and Tschudi}(1981)}]{leyvraz1981}
\bibinfo{author}{\bibfnamefont{F.}~\bibnamefont{Leyvraz}} \bibnamefont{and}
  \bibinfo{author}{\bibfnamefont{H.}~\bibnamefont{Tschudi}},
  \bibinfo{journal}{J. Phys. A} \textbf{\bibinfo{volume}{14}},
  \bibinfo{pages}{3389} (\bibinfo{year}{1981}).

\bibitem[{\citenamefont{Lushnikov}(1978)}]{lushnikov1978}
\bibinfo{author}{\bibfnamefont{A.}~\bibnamefont{Lushnikov}},
  \bibinfo{journal}{Izv. Akad. Nauk SSSR, Ser. Fiz. Atmosfer. I Okeana}
  \textbf{\bibinfo{volume}{14}}, \bibinfo{pages}{738} (\bibinfo{year}{1978}).

\bibitem[{\citenamefont{Ziff}(1980)}]{ziff1980}
\bibinfo{author}{\bibfnamefont{R.}~\bibnamefont{Ziff}}, \bibinfo{journal}{J.
  Stat. Phys.} \textbf{\bibinfo{volume}{23}}, \bibinfo{pages}{241}
  (\bibinfo{year}{1980}).

\bibitem[{\citenamefont{Leyvraz}(2003)}]{leyvraz2003}
\bibinfo{author}{\bibfnamefont{F.}~\bibnamefont{Leyvraz}},
  \bibinfo{journal}{Phys. Reports} \textbf{\bibinfo{volume}{383}},
  \bibinfo{pages}{95} (\bibinfo{year}{2003}).

\bibitem[{\citenamefont{Zaboronski}(2001)}]{zaboronski2001}
\bibinfo{author}{\bibfnamefont{O.}~\bibnamefont{Zaboronski}},
  \bibinfo{journal}{Phys. Lett. A} \textbf{\bibinfo{volume}{281}},
  \bibinfo{pages}{119} (\bibinfo{year}{2001}).

\bibitem[{\citenamefont{Krishnamurthy et~al.}(2002)\citenamefont{Krishnamurthy,
  Rajesh, and Zaboronski}}]{krishnamurthy2002}
\bibinfo{author}{\bibfnamefont{S.}~\bibnamefont{Krishnamurthy}},
  \bibinfo{author}{\bibfnamefont{R.}~\bibnamefont{Rajesh}}, \bibnamefont{and}
  \bibinfo{author}{\bibfnamefont{O.}~\bibnamefont{Zaboronski}},
  \bibinfo{journal}{Phys. Rev. E} \textbf{\bibinfo{volume}{66}},
  \bibinfo{pages}{066118} (\bibinfo{year}{2002}).

\bibitem[{\citenamefont{Connaughton et~al.}(2004)\citenamefont{Connaughton,
  Rajesh, and Zaboronski}}]{cardy2004}
\bibinfo{author}{\bibfnamefont{C.}~\bibnamefont{Connaughton}},
  \bibinfo{author}{\bibfnamefont{R.}~\bibnamefont{Rajesh}}, \bibnamefont{and}
  \bibinfo{author}{\bibfnamefont{O.}~\bibnamefont{Zaboronski}},
  \emph{\bibinfo{title}{Kolmogorov spectra, intermittency and multifractality
  in cluster-cluster aggregation}}, \bibinfo{howpublished}{unpublished}
  (\bibinfo{year}{2004}).

\bibitem[{\citenamefont{Kontorovich}(2001)}]{kontorovich2001}
\bibinfo{author}{\bibfnamefont{V.}~\bibnamefont{Kontorovich}},
  \bibinfo{journal}{Physica D} \textbf{\bibinfo{volume}{152-153}},
  \bibinfo{pages}{676} (\bibinfo{year}{2001}).

\end{thebibliography}

\end{document}